\documentclass[12pt]{iopart}
\usepackage{iopams}
\usepackage{graphicx}
\usepackage{amssymb,amsthm}

\usepackage{graphicx}
\usepackage{bm}
\usepackage{enumerate} 
\begin{document}
%%%%%%%%%%%%%%%%%%%%%%%%%%%%%%%%%%%%%%%%%%%%%%%%%%%%%%%%%%%%
\title{A new method to sum divergent power series: educated match}

\author{Gabriel \'Alvarez}
\address{Departamento de F\'{\i}sica Te\'orica II,
                               Facultad de Ciencias F\'{\i}sicas,
                               Universidad Complutense,
                               28040 Madrid, Spain}
\ead{galvarez@fis.ucm.es}
\author{Harris J. Silverstone}
\address{ Department of Chemistry,
                The Johns Hopkins University,
                3400 N. Charles Street,
                Baltimore, Maryland 21218, USA}
\ead{hjsilverstone@jhu.edu}
%%%%%%%%%%%%%%%%%%%%%%%%%%%%%%%%%%%%%%%%%%%%%%%%%%%%%%%%%%%%
\begin{abstract}
We present a method to sum Borel- and Gevrey-summable asymptotic series by matching the series
to be summed with a linear combination of asymptotic series of known functions that themselves are scaled versions
of a single, appropriate, but otherwise unrestricted, function $\Phi$.  Both the scaling and
linear coefficients are calculated from Pad\'e approximants of a series 
transformed from the original series by $\Phi$. We discuss in particular
the case that $\Phi$ is (essentially) a confluent hypergeometric function, which includes as
special cases the standard Borel-Pad\'e and Borel-Leroy-Pad\'e methods. A particular advantage is the mechanism to
build knowledge about the summed function into
the approximants, extending their accuracy and range even
when only a few coefficients are available. 
Several examples from field theory and Rayleigh-Schr\"odinger perturbation theory illustrate the method.
\end{abstract}
%%%%%%%%%%%%%%%%%%%%%%%%%%%%%%%%%%%%%%%%%%%%%%%%%%%%%%%%%%%%
\maketitle
%%%%%%%%%%%%%%%%%%%%%%%%%%%%%%%%%%%%%%%%%%%%%%%%%%%%%%%%%%%%
%% INTRODUCTION %%%%%%%%%%%%%%%%%%%%%%%%%%%%%%%%%%%%%%%%%%%%%%%%%
%%%%%%%%%%%%%%%%%%%%%%%%%%%%%%%%%%%%%%%%%%%%%%%%%%%%%%%%%%%%
\section{Introduction}
%%%%%%%%%%%%%%%%%%%%%%%%%%%%%%%%%%%%%%%%%%%%%%%%%%%%%%%%%%%% 
%%%%%%%%%%%%%%%%%%%%%%%%%%%%%%%%%%%%%%%%%%%%%%%%%%%%%%%%%%%%
Summation of divergent asymptotic expansions has led to a vast literature from both mathematical and
physical points of view. The mathematical goal is often to assign a standard sum to a series whose coefficients satisfy 
certain growth conditions and whose sum satisfies certain conditions at infinity~\cite{HA49,RA93}.
The physical literature focuses on a wide range of specialized, computational methods. Especially since the work
carried out in the 1970's on the coupling constant analyticity of anharmonic oscillators~\cite{GR70,SI70,HS78},
two summation methods have become dominant: Pad\'e approximants and
Borel summation. Both have been found useful in fields as diverse as quantum mechanics, statistical mechanics,
quantum field theory, and string theory. Pad\'e approximants are most often directly used empirically
(see, for example, the recent study on the existence of an ultraviolet zero for the six-loop beta function
of the $\lambda\Phi_4^4$ theory~\cite{SH16}), and at times with new, alternative transformation procedures~\cite{AM07}.
Borel summability has been rigorously proved in several instances. The analytic continuation implicit
in the Borel summation process poses a practical problem that has been dealt with in essentially two ways:
conformal mappings~\cite{LG77,ZJ02,ZJ10}, and Borel-Pad\'e approximants. In the latter, the analytic
continuation is again performed empirically by Pad\'e approximants
of the Borel-transformed series~\cite{GR70,BA76,FR85,AL00}.
Most recently, Mera, Pedersen, and Nikoli\'c~\cite{ME15,MH16} and Pedersen, Mera, and Nikoli\'c~\cite{PE16}
have developed a method that uses hypergeometric functions
to sum perturbation theory series using only a few terms. 

Initially motivated in part by the papers of Mera, Pedersen, and Nikoli\'c, we present
here a new method to build concise, explicit, analytic approximants to the Borel or Gevrey sum of an asymptotic power
series. These approximants {\em match} the series to be summed with a linear combination of asymptotic series of known functions.
The known functions are scaled versions of a single function $\Phi$, and both the scaling and linear coefficients are readily calculated from Pad\'e approximants of 
a transformed series determined by the original series and by $\Phi$. If $\Phi$ is taken to be
(essentially) a confluent hypergeometric function, the new method includes as
special cases the standard Borel-Pad\'e and Borel-Leroy-Pad\'e summation methods.
Even more important, prior additional (i.e., {\em educated})  knowledge about the summed function can be built into
the approximants via the function $\Phi$, sometimes dramatically extending the accuracy and range of the approximants.
The ``linear combination'' here is similar to the linear combination of the Janke-Kleinert resummation algorithm,
which is described as ``re-expanding the asymptotic expansion in a complete set of basis functions'',
and which is mathematically equivalent to conformal mapping techniques~\cite{KL01}. Our method, in contrast, is essentially linked to the theory of Pad\'e approximants.
%%%%%%%%%%%%%%%%%%%%%%%%%%%%%%%%%%%%%%%%%%%%%%%%%%%%%%%%%%%%
%% THE APPROXIMANTS  %%%%%%%%%%%%%%%%%%%%%%%%%%%%%%%%%%%%%%%%%%%%%%
%%%%%%%%%%%%%%%%%%%%%%%%%%%%%%%%%%%%%%%%%%%%%%%%%%%%%%%%%%%%
\section{$\Phi$-Pad\'e approximants}
%%%%%%%%%%%%%%%%%%%%%%%%%%%%%%%%%%%%%%%%%%%%%%%%%%%%%%%%%%%%
%%%%%%%%%%%%%%%%%%%%%%%%%%%%%%%%%%%%%%%%%%%%%%%%%%%%%%%%%%%%
Our goal is to approximate the Borel sum $\psi(z)$ of a divergent power series,
\begin{equation}
	\label{eq:psi}
	\psi(z) \sim \sum_{k=0}^\infty d_k z^k ,
\end{equation}
using {\em any appropriate known function} $\Phi(z)$ with its own Borel-summable series,
\begin{equation}
	\Phi(z) \sim \sum_{k=0}^\infty f_k (-z)^k.
	\label{eq:Phi}
\end{equation}
The method is at the same time hidden in, and a generalization of, the Borel-Pad\'e summation
method~\cite{AL00}, which we briefly review.

Let us denote by $P_{n-1}(z)/Q_n(z)$ the $[n-1,n]$ Pad\'e approximant of the Borel transform
of the series~(\ref{eq:psi}),
\begin{equation}
	\hat{\psi}_\mathrm{B}(z) = \sum_{k=0}^\infty \frac{d_k}{k!} z^k,
\end{equation}
and let us assume that $Q_n(z)$ has only simple zeros. Partial fraction expansion,
\begin{equation}
	\frac{P_{n-1}(z)}{Q_n(z)} = \sum_{j=1}^n \frac{r_j}{z-z_j},
\end{equation}
and term-by-term integration lead to the standard Borel-Pad\'e approximant
$\psi_{\mathrm{B},[n-1,n]}(z)$ to $\psi(z)$,
\begin{eqnarray}
	\psi_{\mathrm{B},[n-1,n]}(z) & = & \int_0^\infty \rme^{-t} \sum_{j=1}^n \frac{r_j}{z t-z_j} \rmd t\\
	                                    & = & \sum_{j=1}^n \frac{r_j}{-z_j} E_\mathrm{Euler}(-z/z_j),
\end{eqnarray}
where we define $E_\mathrm{Euler}(z)$ by
\begin{equation}
	\label{eq:euler}
	E_\mathrm{Euler}(z) = \int_0^\infty \frac{\rme^{-t}}{1+zt} \rmd t
	                          = z^{-1} \rme^{1/z} E_1(1/z) ,
\end{equation}
and where $E_1(1/z)$ is a standard version of the exponential integral (see chapter~5 of reference~\cite{AS70}).

%Novi Commentarii academiae scientiarum Petropolitanae 5, 1760, pp. 205-237

The two points to note in this derivation are (i) that the $E_\mathrm{Euler}(z)$ in equation~(\ref{eq:euler})
is precisely the Borel sum of the factorially divergent Euler series~\cite{EL1760} obtained by setting $f_k = k!$
in equation~(\ref{eq:Phi}), and (ii) that the asymptotic expansion of $\psi_{\mathrm{B},[n-1,n]}(z)$ is identical
to that of $\psi(z)$ through order $z^{2n-1}$, i.e., that
\begin{equation}
 	\sum_{j=1}^n \frac{r_j}{-z_j} E_\mathrm{Euler} (-z/z_j)
	 = \sum_{k=0}^{2n-1} d_k z^k + O(z^{2n}) .
	\label{eq:reexpansion}
\end{equation}
In principle, the $2n$ parameters $z_j$ and $r_j$ could have been determined \emph{de nouveau}
from equation~(\ref{eq:reexpansion}) by substituting in it the Euler series and equating coefficients.

These observations motivate the following generalization of the Borel-Pad\'e summation method.
We define the ``$\Phi$-transform'' of the series given in equation~(\ref{eq:psi}) by
\begin{equation}
	\label{eq:phitrans}
	\hat{\psi}_\Phi(z) = \sum_{k=0}^\infty \frac{d_k}{f_k} z^k,
\end{equation}
and the associated new approximants $\psi_{\Phi,[n-1,n]}(z)$ to $\psi(z)$ by
\begin{equation}
	\label{eq:na}
	 \psi_{\Phi,[n-1,n]}(z) = \sum_{j=1}^n  \frac{r_j}{-z_j}  \Phi (-z/z_j).
\end{equation}	
As a generalization of equation~(\ref{eq:reexpansion}), the approximants $ \psi_{\Phi,[n-1,n]}(z)$,
which depend on $2n$ parameters $r_j, z_j, (j=1,\dots,n)$, satisfy
\begin{equation}
	 \sum_{j=1}^n  \frac{r_j}{-z_j}  \Phi (-z/z_j)
	 =
	 \sum_{k=0}^{2n-1} d_k z^k +O(z^{2n}),
\end{equation}
and therefore the $r_j$ and $z_j$ solve the $2n$ equations
\begin{equation}
	\sum_{j=1}^n \frac{r_j}{-z_j} f_k ( z_j )^{-k} = d_k , \quad(k=0,1,\ldots,2n-1).
\label{eq:match}
\end{equation}
In practice, these parameters are most easily calculated from the partial fraction expansion
of the $[n-1,n]$ Pad\'e approximant to $\hat{\psi}_\Phi(z)$, i.e., 
\begin{equation}
	\label{eq:PQ}
	\frac{P_{n-1}(z)}{Q_n(z)} = \sum_{k=0}^{2n-1} \frac{d_k}{f_k}z^k + O(z^{2n})
	=
	\sum_{j=1}^n \frac{r_j}{z-z_j}.
\end{equation}
In other words, the $z_j$ are the poles, for simplicity assumed to be simple, and the $r_j$ the residues,
of the $[n-1,n]$ Pad\'e approximant to $\hat{\psi}_\Phi(z)$. Accordingly we call $ \psi_{\Phi,[n-1,n]}(z)$ the
``$[n-1,n]\ \Phi$-Pad\'e approximant'' to $\psi(z)$.

The Borel-Pad\'e approximant uses no information about the sum $\psi(z)$ except for Borel summability.
Generally these approximations will not be accurate over the full range of the variable $z$.
By  an ``educated'' choice of $\Phi(z)$, we mean building additional knowledge about the nature of $\psi(z)$
into $\Phi(z)$, which may lead to very accurate approximations over the full range of the variable $z$
even when only a very limited number of coefficients $d_k$ of the original asymptotic series are available.
Typical examples of prior knowledge that can be built into the $\Phi$-Pad\'e approximations are the
large $z$ behavior of $\psi(z)$ or perhaps the large $k$ behavior of the coefficients $d_k$.
%%%%%%%%%%%%%%%%%%%%%%%%%%%%%%%%%%%%%%%%%%%%%%%%%%%%%%%%%%%%
\subsection{The confluent hypergeometric $\Phi$}
%%%%%%%%%%%%%%%%%%%%%%%%%%%%%%%%%%%%%%%%%%%%%%%%%%%%%%%%%%%%
A prime candidate for $\Phi$ is the confluent hypergeometric function $U$ (see chapter~13 of reference~\cite{AS70})
or, more precisely, the function
\begin{equation}
	\label{eq:hgu}
	\Phi(z) = z^{-a}U \left( a,1+a-b, 1/z \right),
\end{equation}
for which the coefficients $f_k$ in equation~(\ref{eq:Phi}) are
\begin{equation}
	\label{eq:fk}
	f_k =  \frac{ (a)_k (b)_k } { k! },
\end{equation}
where the Pochhammer symbol $(c)_k$ is defined by $(c)_k = \Gamma(c+k)/\Gamma(c)$.
Note that this $\Phi(z)$ is symmetric in $a$ and $b$, which is more obvious from equation~(\ref{eq:fk}) than from equation~(\ref{eq:hgu}).
From a theoretical point of view the confluent hypergeometric $U$ is a natural choice for at least two reasons.
(i) the Borel-Pad\'e method is the special case $a=b=1$, since
\begin{equation}
	z^{-1} U\left( 1,1, 1/z \right) =  z^{-1} \rme^{1/z} E_1(1/z) ,
\end{equation}
which is the $E_\mathrm{Euler}(z)$ of equation~(\ref{eq:euler}).
(ii) Just as the Borel transform is inverted by the Laplace transform, there is a generalization
(which we state without proof) that inverts the ``confluent hypergeometric transform''
[see equations~(\ref{eq:phitrans}) and~(\ref{eq:fk})]: if
\begin{equation}
  \hat{\psi}_\Phi(z) = \sum_{k=0}^\infty \frac { d_k k! } {(a)_k (b)_k}  z^k,
\end{equation}
then
\begin{equation}
    \psi(z)  = \frac{1}{ \Gamma(a)  \Gamma(b) }
                   \int _0^\infty 
                   \hat{\psi}_\Phi( z s ) 
                   \rme^{-s} s^{a-1} U( 1 - b , a - b + 1 , s ) \rmd s.
\label{eq:inverseU}
\end{equation}
(When $b=1$, $U( 0 , a  , t ) = 1$, and the result is the Borel-Leroy transformation~\cite{ZJ10}.)
From a practical point of view, the confluent hypergeometric function~(\ref{eq:hgu}) is also a very
convenient choice, because as $z\to\infty$,
\begin{eqnarray}
	\Phi(z) & \sim & z^{-b} \frac{ \Gamma (a-b)}{\Gamma (a)}+ z^{-a} \frac{ \Gamma (b-a)}{\Gamma (b)},
  		\quad (a-b\ne \mathrm{integer}),
		 \label{eq:abLargeg} \\
		& \sim  & z^{-a}  \frac{  \log ( z ) - 2 \gamma -\psi ^{(0)} ( a ) }{\Gamma (a)},
		\quad (a=b),
\label{eq:aaLargeg}
\end{eqnarray}
where $\gamma$ is Euler's constant and $\psi^{(0)}(a)$ is the polygamma function.
Since the approximant $\psi_{\Phi,[n-1,n]}(z)$ depends linearly on $\Phi$ [see equation~(\ref{eq:na})],
an appropriate choice of $a$ and $b$ permits the large $z$ behavior (if known) of $\psi(z)$ to be built into the $\Phi$-Pad\'e
approximants. We illustrate these general ideas with several examples and generalizations of the method.
%%%%%%%%%%%%%%%%%%%%%%%%%%%%%%%%%%%%%%%%%%%%%%%%%%%%%%%%%%%%
%% EXAMPLES %%%%% %%%%%%%%%%%%%%%%%%%%%%%%%%%%%%%%%%%%%%%%%%%%%%
%%%%%%%%%%%%%%%%%%%%%%%%%%%%%%%%%%%%%%%%%%%%%%%%%%%%%%%%%%%%
\section{Examples}
%%%%%%%%%%%%%%%%%%%%%%%%%%%%%%%%%%%%%%%%%%%%%%%%%%%%%%%%%%%%
%%%%%%%%%%%%%%%%%%%%%%%%%%%%%%%%%%%%%%%%%%%%%%%%%%%%%%%%%%%%
%%%%%%%%%%%%%%%%%%%%%%%%%%%%%%%%%%%%%%%%%%%%%%%%%%%%%%%%%%%%
\subsection{Zero-dimensional $\phi^4$ field theroy}
%%%%%%%%%%%%%%%%%%%%%%%%%%%%%%%%%%%%%%%%%%%%%%%%%%%%%%%%%%%%
As the simplest example, the confluent hypergeometric $\Phi=(\frac{3}{2g})^{3/4}U\left( \frac{3}{4},\frac{3}{2},\frac{3}{2g}  \right)$ trivially sums the perturbative series
for the partition function $Z(g)$ of zero-dimensional $\phi^4$ theory~\cite{ZJ02,ZJ10},
because
\begin{eqnarray}
	Z(g) & = & \frac{1}{\sqrt{2\pi}}\int_{-\infty}^{\infty} \rme^{-x^2/2-gx^4/4!} \rmd x \\
	       & = & \left(  {3}/({2g}) \right)^{3/4}U \left({3}/{4}, {3}/{2}, {3}/({2g})  \right)
\end{eqnarray}
is equal to the $\Phi$ of equation~(\ref{eq:hgu}) with $a=3/4$, $b=1/4$ and $z=2g/3$.
In fact, the $[0,1]$ approximant to the asymptotic expansion of $Z(g)$,
\begin{equation}
	Z(g)  \sim 
	              \sum _{k=0}^\infty
	               \frac{ \Gamma \left(k+\frac{3}{4}\right) \Gamma \left(k+\frac{1}{4}\right)}{\Gamma \left(\frac{3}{4}\right) \Gamma \left(\frac{1}{4}\right) k!}
	               \left(-\frac{2 g}{3}\right)^k
\end{equation}
has $z_1=-3/2$, $r_1=3/2$, and is exactly $Z(g)$.
%%%%%%%%%%%%%%%%%%%%%%%%%%%%%%%%%%%%%%%%%%%%%%%%%%%%%%%%%%%%
\subsection{The Euler-Heisenberg effective Lagrangian}
%%%%%%%%%%%%%%%%%%%%%%%%%%%%%%%%%%%%%%%%%%%%%%%%%%%%%%%%%%%%
A second physically relevant example 
is the Euler-Heisenberg effective Lagrangian~\cite{HE36,DU04}. For the spinor case
in a purely magnetic background,
\begin{equation}
	\label{eq:bpl}
	\mathcal{L}(g) = \int_0^\infty{\rme}^{-s/g} \left(\coth s - \frac{1}{s} - \frac{s}{3}\right)\frac{\rmd s}{s^2},
\end{equation}
(cf.~equations~(1.18) and~(1.19) in reference~\cite{DU04}), and has the asymptotic expansion,
\begin{eqnarray}
	\label{eq:ehs}
	\mathcal{L}(g) & \sim & \sum_{k=0}^\infty \frac{B_{2k+4}}{(2k+4)(2k+3)(2k+2)}(2g)^{2k+2}
\\
	\label{eq:ehss}
	 & \sim & -\frac{1}{45}g^2 + \frac{4}{315} g^4 -\frac{8}{315} g^6 +\cdots,
\end{eqnarray}
where $B_{2k+4}$ denote Bernouilli numbers.
Standard Borel-Pad\'e summation of equation~(\ref{eq:ehs}) would involve Pad\'e approximants in $g^2$ that lead to rational functions of $s^2$, i.e., {\em even} functions of $s$, that have to approximate the Borel transform, which is an {\em odd} function of $s$ (essentially the non-exponential factor in the integrand of equation~(\ref{eq:bpl})). 
This parity clash can be avoided by taking 
\begin{equation}
	\label{eq:EH21}
	\Phi( z ) = z^{-2}U(2,2,1/z),
\end{equation}
i.e., $a=2$, $b=1$, and $f_k=(k+1)!$ rather than $k!$.
The inverse confluent hypergeometric transform equation~(\ref{eq:inverseU}) contains the explicit factor $s$, so that the $\Phi$-transform
with $a=2$ and $b=1$ is in fact an even function of $s$:
\begin{equation}
	\label{eq:L21}
	\hat{\mathcal{L}}_{ \Phi,a=2,b=1} (s) = \left(\coth s - \frac{1}{s} - \frac{s}{3}\right)\frac{1}{s^3}.
\end{equation}
For every $n\ge 1$, all the poles $z_j, (j=1,2,\ldots,n),$ of the $[n-1,n]$ Pad\'e approximants in $s^2$ to $\hat{\mathcal{L}}_{ \Phi,a=2,b=1} (s)$ are negative and simple, meaning that the poles in $s$ are paired on the imaginary axis. The resulting approximants have the form,
\begin{equation}
  \mathcal{L}_{ \Phi,a=2,b=1;[n-1,n]} (g) = 
  \sum_{j=1}^n
  \frac{r_j}{-z_j}
   \frac{1}{2}
   \left( 
   \Phi(\rmi g/\sqrt{-z_j}) +  \Phi(-\rmi g/\sqrt{-z_j})
  \right) ,
\label{eq:EHG1}
\end{equation}
with the $\Phi(z)$ given by equation~(\ref{eq:EH21}). For example, the first Pad\'e approximant to the $\Phi$-transformed series is
\begin{equation}
	-\frac{g^2}{45} \frac{21}{2} \frac{1}{g^2+\frac{21}{2} } 
	 \sim
	-\frac{g^2}{45}
	\left(
	1 - 45\frac{ 4}{315}\frac{1}{3!}g^2 +\cdots
	\right) ,
\end{equation}
with $z_1=-\frac{21}{2}$,  $r_1=-\frac{g^2}{45}\frac{21}{2}$, and the corresponding $\Phi$-Pad\'e approximant is
\begin{equation}
 \mathcal{L}_{ \Phi,a=2,b=1;[0,1]} (g) = 
	-\frac{g^2}{45}
	 \frac{1}{2}
   \left( 
   \Phi ( \rmi g/\sqrt{21/2} ) +  \Phi ( -\rmi g/\sqrt{21/2} )
  \right).
\end{equation}
If expanded as a power series in $g$, this simple approximation reproduces
the first two nonvanishing terms of equation~(\ref{eq:ehss}), but at the same time it also captures
the functional form of the large-$g$ expansion: in fact $\mathcal{L}_{ \Phi,a=2,b=1;[0,1]} (g) \sim  -(7/30)\log(g)$,
while the exact result is $\mathcal{L}(g)\sim  -(1/3)\log(g)$~\cite{DU04}. Note that
the exact expansion,
\begin{equation}
	\left(\coth s - \frac{1}{s} - \frac{s}{3}\right)\frac{1}{s^3}
	=
	 \sum_{j=1}^\infty \frac{-2}{ j^2 \pi^2 ( j^2 \pi^2 +s^2)} ,
\end{equation}
can be viewed as the  ``$[\infty-1,\infty]$'' Pad\'e approximant in $s^2$ for the $\Phi$-transform,
from which the exact poles and residues can be read off:
\begin{equation}
	z_j = - j^2 \pi^2,
\quad
	r_j  = -\frac{2}{ j^2 \pi^2 }.
\end{equation}
With $\Phi$ given by equation~(\ref{eq:EH21}), the resulting $\Phi$-Pad\'e infinite sum reproduces $\mathcal{L}(g)$:
\begin{equation}
	\mathcal{L}_{ \Phi,a=2,b=1;[\infty-1,\infty]} (g) = 
	\sum_{j=1}^\infty
	\frac{ - 2 }{ j^4 \pi^4 }
	\frac{ 
	\Phi( \rmi g / ( j \pi )) +  \Phi( - \rmi g / ( j \pi ))}
	{2}.
\end{equation}
We remark in passing that the coefficients $-2/(j^{4}\pi^{4})$ give the rate of convergence of the approximants.
%%%%%%%%%%%%%%%%%%%%%%%%%%%%%%%%%%%%%%%%%%%%%%%%%%%%%%%%%%%%
\subsection{One-dimensional $\phi^4$ field theory: the quartic anharmonic oscillator}
\begin{figure}[htb]
	\begin{center}
		\includegraphics[scale=0.95]{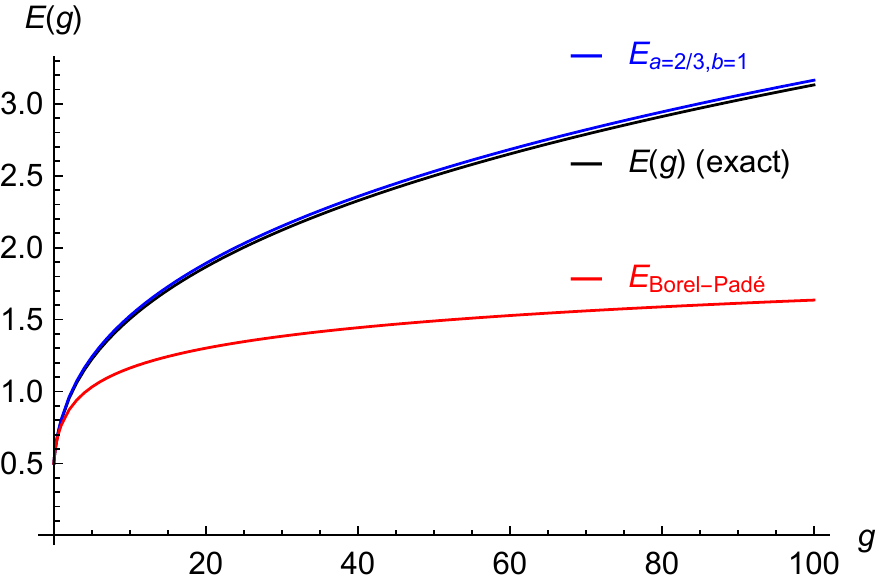}
	\end{center}
	\caption{Exact $E(g)$ (black) and [0,1] approximants for Borel-Pad\'e (red) and
	              $(a=2/3,b=1)$ confluent hypergeometric $\Phi$ (blue). The confluent hypergeometric $\Phi$
	              approximant agrees well with the exact $E(g)$, because the $g^{1/3}$ large-$g$
	              behavior is carried by the $\Phi(g)$.}
	\label{fig:1}
\end{figure}

%%%%%%%%%%%%%%%%%%%%%%%%%%%%%%%%%%%%%%%%%%%%%%%%%%%%%%%%%%%%
Third, we consider one-dimensional $\phi^4$ theory, i.e., the familiar $x^4$-perturbed
anharmonic oscillator, whose Schr\"odinger equation is
\begin{equation}
	\label{eq:x4sch}
  \left(-\frac{1}{2}\frac{d^2}{dx^2} +\frac{1}{2}x^2 +gx^4\right)\Psi(x) = E(g) \Psi(x).
  \end{equation}
The first three coefficients of the ground state Rayleigh-Schr\"odinger perturbation series are
 \begin{equation}
 	E(g) = \frac{1}{2}+\frac{3}{4} g -\frac{21}{8}  g^2 + \cdots.
	\label{eq:x4RSPT}
\end{equation}
The coefficients $E^{(k)}$ of this Borel-summable~\cite{GR70} series behave like
\begin{equation}
   \label{eq:anhoCoef}
	E^{(k)} \sim 
	(-1)^{k+1} \frac{2^{1/2} 3^{k+\frac{1}{2}}}{ \pi^{3/2}} \Gamma\left(k+\frac{1}{2}\right),
	\quad k\to\infty .
\end{equation}
More important is the large-$g$ behavior of $E(g)$, which follows from a simple scaling argument, 
\begin{equation}
	E(g) \sim g^{1/3} \varepsilon, \quad \mathrm{as }\quad g\rightarrow \infty,
\end{equation}
where $\varepsilon = 0.667986\ldots$ is the ground state energy of the purely quartic oscillator.
If the $g^{1/3}$ behavior is built into  $\Phi$, then even a two-parameter [0,1] approximant gives an excellent fit to $E(g)$ all the way from 0 to $\infty$.
The details are elementary enough to execute by hand. Because of the sign pattern, we sum the once-subtracted
series,
\begin{equation}
	\psi(g) = \frac{E(g)-1/2}{g},
\end{equation}
whose large-$g$ behavior is $g^{-2/3}$ (then multiply by $g$ and add 1/2 to report the results). 
Equation~(\ref{eq:abLargeg}) shows that a suitable $\Phi$ with this behavior can be obtained by taking
$a=2/3$ and $b>a$ in equation~(\ref{eq:hgu}). If $b$ were then chosen to fit the exact quartic $\varepsilon$, its value would be $0.9977547\ldots$.
We take $b=1$ (Borel-Leroy-Pad\'e, but note that $a=2/3$ is different from that implied by equation~(\ref{eq:anhoCoef})). The $[0,1]$ Pad\'e approximant to the transformed series, which needs only the two coefficients $3/4$ and $-21/8$ from the $E(g)$-series and $f_1=2/3$ from the $\Phi$-series, has $z_1=-4/21$ and $r_1=1/7$. The [0,1] $\Phi$-Pad\'e approximant is
\begin{equation}
	\label{eq:x4blp}
	   E_{\Phi,[0,1]}(g) = \frac{1}{2} + 
	\frac{3}{4} 
	\left(   \frac {4} {21}   \right)^{2/3}
	g^{1/3}
	U \left( \frac{2}{3},\frac{2}{3}, \frac{4}{21g} \right),
\end{equation}
which, despite its simple origins, turns out to give remarkable agreement with $E(g)$ for all $g>0$,
as seen in Fig.~\ref{fig:1}. At $\infty$,  
\begin{eqnarray}
   E_{\Phi,[0,1]}(g) & \sim &
	\frac{3}{4} 
	\left(   \frac {4} {21}   \right)^{2/3}
	\Gamma\left( \frac{1}{3}\right)  
		g^{1/3}
\\
	& = &  0.665147\ldots g^{1/3},\quad\mathrm{as }\quad g\rightarrow \infty ;
\end{eqnarray}
the constant $0.665147\ldots$ is within 0.4\% of the exact quartic $\varepsilon$.
Higher-order $[n-1,n]$
approximants generally agree progressively better. It is clear from Fig.~\ref{fig:1} in which the [0,1] Borel-Pad\'e approximant is also plotted, 
how relatively simple information used to choose the function generating the match
can dramatically affect the quality and range of the approximant.\footnote{All numerical calculations have been done
in extended precision using {\em Mathematica}, version 11.1; the commands, PadeApproximant and HypergeometricU, were particularly relevant.}

%%%%%%%%%%%%%%%%%%%%%%%%%%%%%%%%%%%%%%%%%%%%%%%%%%%%%%%%%%%%
%%%%%%%%%%%%%%%%%%%%%%%%%%%%%%%%%%%%%%%%%%%%%%%%%%%%%%%%%%%%
%%%%%%%%%%%%%%%%%%%%%%%%%%%%%%%%%%%%%%%%%%%%%%%%%%%%%%%%%%%%
\subsection{Implementation of the large-order behavior of the perturbation coefficients}
%%%%%%%%%%%%%%%%%%%%%%%%%%%%%%%%%%%%%%%%%%%%%%%%%%%%%%%%%%%%
As an example of the versatility of the method we show how to incorporate in a simple way the asymptotic
behavior of the coefficients $d_k$ into the function $\Phi$. We consider the $\beta$-function for the $\phi^4$ theory in
$d=3$ dimensions~\cite{ZJ10}, with coefficients
\begin{eqnarray}
	\tilde{\beta}(\tilde{g}) & = 0 - \tilde{g} + \tilde{g}^2 - \frac{308}{729} \tilde{g}^3
		+ 0.3510695977 \tilde{g}^4 
\nonumber \\
		& \quad  - 0.3765268283  \tilde{g}^5  + 0.49554751   \tilde{g}^6
                               - 0.749689   \tilde{g}^7   + O(\tilde{g}^8)
\end{eqnarray}
and growth
\begin{equation}
	\label{eq:betak}
	\tilde{\beta}_k
	\sim (-0.147774232\ldots)^k k^{7/2} k!,
	\quad
	k\to\infty.
\end{equation}
The $[3,4]$ Pad\'e approximant for the Borel transform of $\tilde{\beta}(\tilde{g})$ has a pole on the positive axis at $\tilde{g}=17.34418$ and consequently fails to be analytic in a strip containing the positive real axis, invalidating a possible $[3,4]$ Borel-Pad\'e approximant.
Stirling's formula shows that asymptotically the $f_k$ in equation~(\ref{eq:fk}) go like
\begin{equation}
	f_k \sim 
	\frac{ k! k^{a+b-2} }{ \Gamma(a)\Gamma(b) }
	\left(1+ \frac{a^2-a+b^2-b +1/6 }{2k} \right)
	,
\quad\mathrm{as }\quad
	k\to\infty,
\end{equation}
so that the growth of the coefficients $\tilde{\beta}_k$ in equation~(\ref{eq:betak}) is matched when
$a+b=11/2$; the $1/k$-term is then minimum when $a=b=11/4$.
With this straightforward choice of $a$ and $b$, and with the corresponding $[3,4]$ approximant
to $\tilde{\beta}(\tilde{g})$, we obtain a value for the nontrivial root of the $\beta$-function
of $\tilde{g}^*=1.4192$, which is close to the value 1.4105 of~\cite{ZJ10}. But we have no estimate of the accuracy of our calculation.
%%%%%%%%%%%%%%%%%%%%%%%%%%%%%%%%%%%%%%%%%%%%%%%%%%%%%%%%%%%%
%%%%%%%%%%%%%%%%%%%%%%%%%%%%%%%%%%%%%%%%%%%%%%%%%%%%%%%%%%%%
\section{$\Phi$-Pad\'e approximants for Gevrey-summable series}
%%%%%%%%%%%%%%%%%%%%%%%%%%%%%%%%%%%%%%%%%%%%%%%%%%%%%%%%%%%%
%%%%%%%%%%%%%%%%%%%%%%%%%%%%%%%%%%%%%%%%%%%%%%%%%%%%%%%%%%%%
Next we adapt the new $\Phi$-Pad\'e approximant method to the cases of summable series
whose coefficients $d_k$  grow like $(mk)!$, where $m=2,3,\ldots$, and which are 
variously known as generalized Borel summable~\cite{GR70}, $m$-summable or Gevrey-$1/m$
summable~\cite{RA93}. The $m=2$ case is useful for summing the $x^6$-perturbed oscillator and the Euler-Heisenberg series~(\ref{eq:ehs}), and
$m=3$ is useful for the $x^8$-perturbed oscillator, etc.
We regard these series in $z$ with $(mk)!$ growth to be series in $z^{1/m}$ with $k!$ growth, but in which the coefficients
of all the fractional powers are 0. By averaging over the $m$-th roots of unity, from a given ($k!$)-$\Phi(z)$~[equation~(\ref{eq:Phi})] we can construct $m$
appropriate ``Gevrey-$1/m$'' summed series $\Phi_\mu^{(1/m)}(z)$, $\mu=0, 1, \ldots,m-1$. 
$\Phi_\mu^{(1/m)}(z)$ has the asymptotic series,
\begin{equation}
   \Phi_\mu^{(1/m)}(z) \sim \sum_{k=0}^\infty  f_{\mu+mk}(-z)^k ,
\end{equation}
and the explicit formula,
\begin{equation}
	\label{eq:gev}
  	\Phi_\mu^{(1/m)}(z)  =
	\frac{ 
	\frac{1}{m}
	 \sum_{j=1}^m 
		\omega_m^{-\mu j}
	 \Phi(-
	 	\omega_m^{\  j}
	\rme^{\pi\rmi/m} z^{1/m} ) }
	{
	(\rme^{\pi\rmi/m} z^{1/m})^{\mu}
	 },
\end{equation}
where $\omega_m=\rme^{2\pi\rmi /m}$. The practical procedural consequence
is that $f_k$ gets replaced by $f_{\mu+mk}$ in equations~(\ref{eq:match}) and (\ref{eq:PQ}).
The question, which $\mu$ is appropriate, is similar to which $a$ and $b$ are appropriate, and the answers depend on which properties, e.g., large $z$, $d_k$ for large $k$, etc., are most appropriate for $\psi$. Moreover, the same Gevrey-$1/m$~$\Phi_\mu^{(1/m)}$ can result from two different Gevrey-1 $\Phi$'s with different $\mu$'s, as illustrated in the next three equations and following remark:
If, for instance, 
\begin{equation}
   \Phi(z) \sim \sum_{k=0}^\infty k!(-z)^k ,
\label{eq:U11}
\end{equation}
then
\begin{eqnarray}
   \Phi_0^{(1/2)}(z) & \sim \sum_{k=0}^\infty (2k)!(-z)^k ,
\\
   \Phi_1^{(1/2)}(z) & \sim \sum_{k=0}^\infty (2k+1)!(-z)^k .
\label{eq:2kp1fact}
\end{eqnarray}
The Euler-Heisenberg integral discussed above, particularly equation~(\ref{eq:EHG1}), is better understood as a Gevrey-1/2 series summed by the $\mu=0$ version of the $\Phi(z)$ given by equation~(\ref{eq:EH21}), which is the same as the $\mu=1$ version of $z^{-1}U(1,1,1/z)$ [equation~(\ref{eq:U11})] given by equation~(\ref{eq:2kp1fact}).
%%%%%%%%%%%%%%%%%%%%%%%%%%%%%%%%%%%%%%%%%%%%%%%%%%%%%%%%%%%%
\subsection{The sextic anharmonic oscillator}
\begin{figure}
	\begin{center}
		\includegraphics[scale=.95]{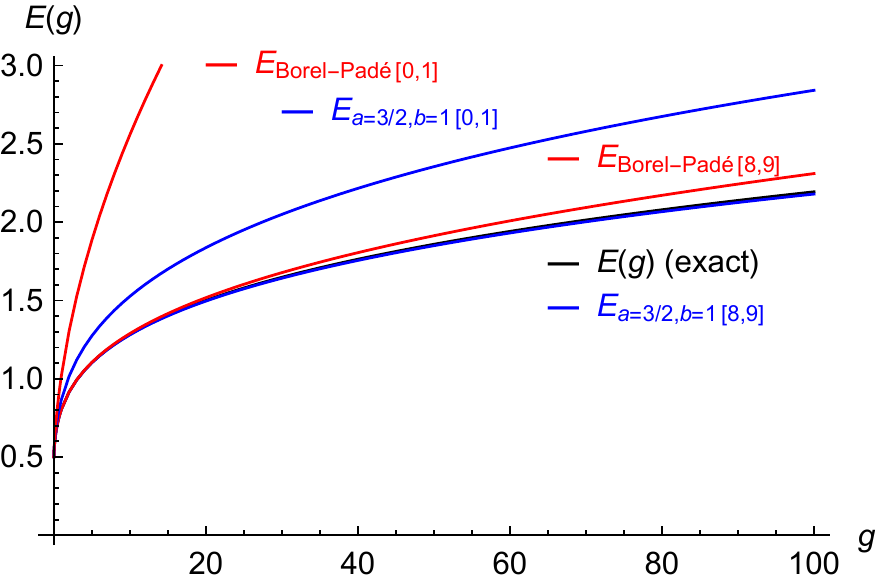}
	\end{center}
	\caption{
		Exact $E(g)$ (black),  Borel-Pad\'e approximants (red), and $(a=3/2,b=1)$-Pad\'e approximants (blue) for the $x^6$-perturbed
		anharmonic oscillator. 
		The $g^{1/4}$ large-$g$ behavior is carried by the  $(a=3/2,b=1)$-confluent-hypergeometric-function-based $\Phi_0^{(1/2)}(g)$.
		The [0,1] and [8,9] approximants are shown. The largest relative error for the $(a=3/2,b=1)$ [8,9] approximant occurs at $g=100$ and is less than 0.007,
		which is barely distinguishable from the exact $E(g)$.}
	\label{fig:2}
\end{figure}

%%%%%%%%%%%%%%%%%%%%%%%%%%%%%%%%%%%%%%%%%%%%%%%%%%%%%%%%%%%%
A classic Gevrey-1/2 series is the Rayleigh-Schr\"odinger perturbation series for the $x^6$-perturbed
anharmonic oscillator (i.e, the Schr\"odinger equation~(\ref{eq:x4sch}) with $g x^4$ replaced by $g x^6$). 
The first three coefficients of the ground-state energy series are
 \begin{equation}
 	E(g) = \frac{1}{2}+\frac{15}{8} g -\frac{3495}{64}  g^2 + \cdots.
\end{equation}
For large $k$, the coefficients $E^{(k)}$ behave like
\begin{equation}
	E^{(k)} \sim 
	(-1)^{k+1} \left(\frac{32}{\pi^2}\right)^{k+1} \Gamma\left(2k+\frac{1}{2}\right),
	\quad k\to\infty ,
\end{equation}
and for large $g$
\begin{equation}
  E(g) \sim g^{1/4} \varepsilon,
\end{equation}
where $\varepsilon$ here is the ground-state energy of the pure $x^6$ oscillator. To build the $g^{1/4}$ behavior into the approximants, we take (for the once-subtracted series) $\Phi(z)=z^{-3/2}U(3/2,1,1/z)$. Although equation~(\ref{eq:abLargeg}) seems to imply that the large-$z$ behavior would be $z^{-1}$ rather than $z^{-3/2}$, the $z^{-1}$ term is canceled in constructing $\Phi_0^{(1/2)}$. When the approximant for the subtracted series is multiplied by $g$, the remaining $(g^{1/2})^{-3/2}$ term gives $g^{1/4}$.
The $[0,1]$ $\Phi$-Pad\'e approximant, which like the $x^4$ case can be done by hand, yields
\begin{equation}
	\label{eq:x601}
	   E_{\Phi,[0,1]}(g) = \frac{1}{2} +  g \frac{15}{8} \Phi_0^{(1/2)}\left(\frac{30}{233 g}\right).
\end{equation}
This simple $[0,1]$ approximation for the sextic oscillator, while superior to the $[0,1]$ Borel-Pad\'e approximant, is not as dramatically accurate as the analogous
approximation for the quartic oscillator given in equation~(\ref{eq:x4blp}), but as $n$ increases
the accuracy of the $[n-1,n]$ $\Phi$-Pad\'e approximant increases monotonically to the point that in Fig.~\ref{fig:2} it is difficult to distinguish
between the exact and [8,9]-approximant values for $0\le g\le 100$. (The maximum relative error at $g=100$ is less than 0.007.) The error in the Borel-Pad\'e approximants is much larger.
%
%%%%%%%%%%%%%%%%%%%%%%%%%%%%%%%%%%%%%%%%%%%%%%%%%%%%%%%%%%%%
%%%%%%%%%%%%%%%%%%%%%%%%%%%%%%%%%%%%%%%%%%%%%%%%%%%%%%%%%%%%
%%%%%%%%%%%%%%%%%%%%%%%%%%%%%%%%%%%%%%%%%%%%%%%%%%%%%%%%%%%%
%%%%%%%%%%%%%%%%%%%%%%%%%%%%%%%%%%%%%%%%%%%%%%%%%%%%%%%%%%%%
\section{Summary}
%%%%%%%%%%%%%%%%%%%%%%%%%%%%%%%%%%%%%%%%%%%%%%%%%%%%%%%%%%%%
%%%%%%%%%%%%%%%%%%%%%%%%%%%%%%%%%%%%%%%%%%%%%%%%%%%%%%%%%%%%
In summary, the conceptualization presented here emphasizes matching the series to be summed
with a linear combination of asymptotic series of known functions, cf. equation~(\ref{eq:na}).
The known functions are scaled versions of a single function $\Phi(z)$, and the scaling 
and  linear coefficients are calculated from the $[n-1,n]$ Pad\'e approximants of the transformed
series generated by $\Phi(z)$. The whole idea stems from the realization that the Borel-Pad\'e approximant has
exactly that structure, but where the $\Phi(z)$ is the sum of Euler's factorially divergent power series,
and from the thought that approximants would be much more accurate if $\Phi(z)$ were more appropriate
for the unknown sum $\psi(z)$. Building the long-range behavior of $\psi$ into $\Phi$ is particularly successful.
%%%%%%%%%%%%%%%%%%%%%%%%%%%%%%%%%%%%%%%%%%%%%%%%%%%%%%%%%%%%
%%%%%%%%%%%%%%%%%%%%%%%%%%%%%%%%%%%%%%%%%%%%%%%%%%%%%%%%%%%%
\section*{Acknowledgments}
We wish to acknowledge the support of the Spanish Ministerio
de Econom\'{\i}a y Competitividad under Project No.~FIS2015-63966-P
 and of the Department of Chemistry of the Johns Hopkins University.
%%%%%%%%%%%%%%%%%%%%%%%%%%%%%%%%%%%%%%%%%%%%%%%%%%%%%%%%%%%%
%%%%%%%%%%%%%%%%%%%%%%%%%%%%%%%%%%%%%%%%%%%%%%%%%%%%%%%%%%%%
%%  BIB %%%%%%%%%%%%%%%%%%%%%%%%%%%%%%%%%%%%%%%%%%%%%%%%%%%%%%%
%%%%%%%%%%%%%%%%%%%%%%%%%%%%%%%%%%%%%%%%%%%%%%%%%%%%%%%%%%%%
\section*{References}
%\bibliographystyle{iopart-num}
%\bibliography{newmethod}

\begin{thebibliography}{10}
\expandafter\ifx\csname url\endcsname\relax
  \def\url#1{{\tt #1}}\fi
\expandafter\ifx\csname urlprefix\endcsname\relax\def\urlprefix{URL }\fi
\providecommand{\eprint}[2][]{\url{#2}}

\bibitem{HA49}
Hardy G~H 1949 {\em Divergent series\/} (Oxford: Clarendon)

\bibitem{RA93}
Ramis J~P 1993 {\em S\'eries Divergentes et Th{\'e}ories Asymptotiques\/} vol
  121 (Marseille: Soci\'et\'e Math\'ematique de France)

\bibitem{GR70}
Graffi S, Grecchi V and Simon B 1970 {\em Phys. Lett. B\/} {\bf 32} 631

\bibitem{SI70}
Simon B 1970 {\em Ann. Phys.\/} {\bf 58} 76

\bibitem{HS78}
Herbst I~W and Simon B 1978 {\em Phys. Lett.\/} {\bf 78B} 304

\bibitem{SH16}
Shrock R 2016 {\em Phys. Rev. D\/} {\bf 94} 125026

\bibitem{AM07}
Amore P 2007 {\em Phys. Rev. D\/} {\bf 76} 076001

\bibitem{LG77}
{Le Guillou} J~C and Zinn-Justin J 1977 {\em Phys. Rev. Lett.\/} {\bf 39} 95

\bibitem{ZJ02}
Zinn-Justin J 2002 {\em Quantum Field Theory and Critical Phenomena\/} (Oxford: Clarendon)

\bibitem{ZJ10}
Zinn-Justin J 2010 {\em Appl. Num. Math.\/} {\bf 60} 1454

\bibitem{BA76}
{Baker Jr} G~A, Nickel B~G, Green M~S and Meiron D~I 1976 {\em Phys. Rev.
  Lett.\/} {\bf 36} 1351

\bibitem{FR85}
Franceschini V, Grecchi V and Silverstone H~J 1985 {\em Phys. Rev. A\/} {\bf
  32} 1338

\bibitem{AL00}
\'Alvarez G, Mart\'{\i}n-Mayor V and Ruiz-Lorenzo J~J 2000 {\em J. Phys.\ A:
  Math.\ Gen.\/} {\bf 33} 841

\bibitem{ME15}
Mera H, Pedersen T~G and Nikoli\'c B~K 2015 {\em Phys. Rev. Lett.\/} {\bf 115}
  143001

\bibitem{MH16}
Mera H, Pedersen T~G and Nikoli\'c B~K 2016 {\em Phys. Rev. B\/} {\bf 94}
  165429

\bibitem{PE16}
Pedersen T~G, Mera H and Nikoli\'c B~K 2016 {\em Phys. Rev. A\/} {\bf 93}
  013409

\bibitem{KL01}
Kleinert H and Schulte-Frohlinde V 2001 {\em Critical properties of
  $\phi^4$-theories\/} (Singapore: World Scientific)

\bibitem{AS70}
Abramowitz M and Stegun I~A (eds) 1970 {\em Handbook of {M}athematical
  {F}unctions\/} (New York: Dover)

\bibitem{EL1760}
Euler L 1760 (1754-55) {\em Novi. Comm. Acad. Sci. Petrop.\/} {\bf 5} 205--237

\bibitem{HE36}
Heisenberg W and Euler H 1936 {\em Z. Phys.\/} {\bf 98} 714

\bibitem{DU04}
Dunne G 2004 Heisenberg-Euler effective Lagrangians: Basics and extensions {\em
  From Fields to Strings: Circumnavigating Theoretical Physics\/} ({\em Ian
  Kogan Memorial Collection\/} vol~I) ed {M A Shifman et al} (Singapore: World Scientific)
  p 445

\end{thebibliography}
%%%%%%%%%%%%%%%%%%%%%%%%%%%%%%%%%%%%%%%%%%%%%%%%%%%%%%%%%%%%
%%%%%%%%%%%%%%%%%%%%%%%%%%%%%%%%%%%%%%%%%%%%%%%%%%%%%%%%%%%%
%%%%%%%%%%%%%%%%%%%%%%%%%%%%%%%%%%%%%%%%%%%%%%%%%%%%%%%%%%%%
\providecommand{\newblock}{}

\end{document}